\documentstyle[amssymb,twocolumn,prb,aps]{revtex}
\large

\begin{document}
\twocolumn[
\hsize\textwidth\columnwidth\hsize\csname @twocolumnfalse\endcsname

\title{Effect of Al doping on the optical phonon spectrum in Mg$_{1-x}$Al$_{x}$B$_{2}$}
\author{P. Postorino (1), A. Congeduti (1), P. Dore (1), A. Nucara (1), 
A. Bianconi (1), D. Di Castro (1), S. De Negri (2), A. Saccone (2)}
\address{$^1$ Dipartimento di Fisica, Universita' di Roma ''La Sapienza'',\\
Unita' dell'Istituto Nazionale per la Fisica per la Materia\\
Piazzale A. Moro 2, 00185 Roma, Italy\\
$^{2}$ Dipartimento di Chimica e Chimica Industriale, Universita' di Genova,\\
Via Dodecanneso 31, 16146 Genova, Italy\\}
\date{\today }
\maketitle

\begin{abstract}
Raman and infrared absorption spectra of Mg$_{1-x}$Al$_{x}$B$_{2}$ have been 
collected for $0\leq x\leq 0.5$ in the spectral range of optical phonons.
The x-dependence of the peak frequency, the width and the intensity of the
observed Raman lines has been carefully analized. A peculiar
x-dependence of the optical modes is pointed out for two different Al doping
ranges. In particular the onset of the high-doping structural phase previously
observed in diffraction measurements is marked by the
appearence of new spectral components at high frequencies. A connection
between the whole of our results and the observed suppression of
superconductivity in the high doping region is established.\\
\end{abstract}

]

The recent discovery\cite{first} of superconductivity below 39 K in MgB$_{2}$
has stimulated a great deal of effort among the scientific community and a
large number of theoretical and experimental papers have been published
within few months. The debate on the origin of this unexpected
superconductivity is still open, although both experimental\cite
{exper1,exper2,exper3} and theoretical \cite{an,kortus,kong} works indicate
that MgB$_{2}$ is a BCS-like system. In this framework, the obvious relevant
interaction in the superconducting transition is the electron-phonon (e-ph)
coupling. Owing to the simple hexagonal structure (space group P$_{6}$mmm),
four zone-center optical modes are predicted for MgB$_{2}$: a silent B$_{1g}$
mode, the E$_{2g}$ Raman mode, and the infrared active E$_{2u}$ and A$_{2u}$
modes. While the doubly-degenerate E$_{2u}$ and E$_{2g}$ modes are ascribed
to in-plane stretching modes of the boron atoms, both non-degenerate 
A$_{2u}$ and B$_{1g}$ modes involve vibrations along the perpendicular
direction (c axis). It is quite a general statement that the E$_{2g}$ mode
is expected to allow for the strongest e-ph coupling\cite{an,kortus,kong} and then to play a
relevant role in superconductivity. Raman experiments\cite
{goncharov,bohnen,hlinka,chen,kunc} carried out on MgB$_{2}$ have shown that
the spectrum is dominated by a quite large and asymmetric band around 600 
cm$^{-1}$, ascribed to the E$_{2g}$ mode. The anomalous width of this phonon
peak has been interpreted as a signature of the e-ph coupling.

Up to now, no other isostructural boride (XB$_{2}$) has shown the peculiar
high temperature superconductivity of MgB$_{2}$. In particular, MgAl$_{2}$ is not
superconducting. Indeed, several studies on the 
Mg$_{1-x}$Al$_{x}$B$_{2}$ compounds have shown that superconductivity is
progressively suppressed for increasing x and
vanishes for x$>$0.5.\cite{slusky,bianconi1,xiang} In order to achieve a
deeper understanding of the effects of Al doping, we have studied the evolution 
of the phonon spectrum of Mg$_{1-x}$Al$_{x}$B$_{2}$ in the $0\leq x\leq 0.5$ 
range by means of both Raman and infrared spectroscopy.

Pure MgB$_{2}$ and Al doped polycrystalline samples have been synthesized at
high temperature by direct reaction of the elements in a tantalum crucible
under argon atmosphere. The samples, which show an average grain dimension
around 1-2$\mu $m, have been characterized by x-ray diffraction and by
resistivity measurements, in order to determine, in particular, the
x-dependence of the superconductivity transition temperature T$_{c}$.\cite{bianconi1,bianconi2}.

The Raman spectra were measured in back-scattering geometry, using a micro-Raman
spectrometer with a CCD detector and an adjustable notch filter. The sample
was excited by the 632.8 nm line of a 16 mW He-Ne Laser. The confocal
microscope was equipped with a 20X magnification objective which gives a
laser spot about 10 $\mu $m$^{2}$ wide at the sample surface. The Raman shift 
explored ranges between 200 and 1100 cm$^{-1}$, the low frequency limit
being due to the notch filter cutoff. Since the laser spot impinges on a few
sample grains, the relative intensities of the observed spectral features
were slightly different from point to point, depending on the random
orientation of the grains. For each sample, Raman spectra were thus collected
from different points and then averaged.

As shown in Fig. 1a, the spectrum of the undoped MgB$_{2}$ sample is
dominated by a band centered around 600 cm$^{-1}$, in agreement with
previous Raman experiments reporting a band centered 
between 580 and 630 cm$^{-1}$, characterized by a large width (200-300 
cm$^{-1} $) and a relevant asymmetry.\cite{goncharov,bohnen,hlinka,chen,kunc} A
close inspection of the MgB$_{2}$ spectrum of Fig. 1a reveals two weak
shoulders on the low- and the high- frequency side of the 600 cm$^{-1}$
band ($\nu _{2}$), around 400 ($\nu _{1}$) and 750 cm$^{-1}$ ($\nu_{3}$),
respectively. Although the origin of the central $\nu _{2}$ band is still
questioned,\cite{chen,kunc} it is generally assigned to the Raman-active
mode E$_{2g}$, to which theoretical predictions attribute a peak frequency\cite
{kortus,kong,bohnen,kunc,satta,yildrim} ranging from 470 to 660 cm$^{-1}$. It is
worth noticing that further spectral contributions observed in previous
Raman experiments\cite{hlinka,kunc} have been ascribed to peaks in the phonon
density of states (around 430, 620, 710 and 780 cm$^{-1}$) derived from neutron scattering 
experiments.\cite{osborn} In a disordered or defective system, the momentum selection 
rules can indeed be violated, and the Raman (or infrared absorption) spectrum can reflect the
phonon density of states.

The Raman spectrum was fitted by a combination of three contributions, each
described by a Damped Harmonic Oscillator (DHO), since this profile has been
successfully used in modelling broad phonon line-shapes in strongly correlated
systems.\cite{dho,prl} The complete fitting function here used is:

\begin{equation}
S(\nu )=[1+n(\nu )]\Bigg[ {\frac{{A\nu \Gamma }}{{\nu ^2+\Gamma ^2}}}%
+\sum_{i=1}^N{\frac{{A_i\nu \Gamma _i}}{{(\nu ^2-\nu _i^2)^2+\nu ^2\Gamma
_i^2}}}\Bigg]
\end{equation}

The first term represents a large DHO function centered at $\nu $=0. It takes
into account the wide unstructured background, which is a common feature
of strongly correlated systems such as manganites and cuprates. The second
term accounts for N phonon peaks, where $\nu _{i}$, A$_{i\text{ }}$and 
$\Gamma_{i} $ are their peak frequency, amplitude and linewidth, respectively. The
quantity n($\nu $) is the Bose-Einstein thermal population factor. In Fig. 1a
we report the best fit spectrum, the $\nu _{1}$, $\nu _{2}$, and $\nu _{3}$
components, and the tail of the background centered at $\nu $=0. As discussed above, the 
$\nu _{2}$ peak (centered at 605$\pm $10\ cm$^{-1}$) can be ascribed to the 
E$_{g} $ mode, while the $\nu _{1}$ (at 415$\pm $20\ cm$^{-1}$) and $\nu _{3}$ (at 
780$\pm $30\ cm$^{-1}$) peaks probably reflect optical bands in the phonon density of
states.\cite{osborn}

The spectra of the doped Mg$_{1-x}$Al$_{x}$B$_{2}$ samples are well
reproduced by using Eq. (1) if, for x$>$0.2, the phonon number N is
increased from 3 to 4. In Fig. 1b we report the Raman spectra for different Al doping after
background subtraction. The $\nu _{3}$\ contribution becomes well
detectable in the x=0.08 spectrum. Both its intensity and peak frequency
increase on further increasing x. For x$>$0.1, a new component 
($\nu _{4}$) appears around 850 cm$^{-1}$ and, for x$>$0.25, the $\nu _{3}$ and 
$\nu _{4}$ peaks become the dominant contributions to the Raman spectrum. It is
interesting to note that the spectrum at x=0.5 resembles that of the
end-series compound (AlB$_{2}$), which is dominated by an intense and narrow
peak centered around 980 cm$^{-1}$.\cite{bohnen}

Infrared absorption measurements were performed above 400 cm$^{-1}$ by using
a Bomem MB100 interferometer operating with a resolution of 10 cm$^{-1}$.
Following a standard procedure,\cite{noilasco} we measured the infrared signals transmitted
by a pure CsI pellet (I$_{o}$($\nu $)) and by the sample powder dispersed in
a CsI pellet (I($\nu $)). The optical density 
$O_{d}(\nu )$=ln(I$_{o}$($\nu $)/I($\nu $)) is proportional to the optical 
conductivity of the sample\cite{noilasco} 
and thus provides the spectral shape of the sample absorption.
When this procedure is employed for measuring a powder metallic sample, the
phonon spectrum is strongly reduced in intensity by the screening from free
charges, and superimposed to a broad and intense background (see for
example the case of metallic La$_{2-x}$Sr$_{x}$CuO$_{4}$ powders\cite{noilasco}). 
The far infrared $O_{d}(\nu )$ of the measured Mg$_{1-x}$Al$_{x}$B$_{2}$ 
are affected by intense backgrounds (see for example the MgB$_{2}$ case in Fig. 2a), 
which prevent reliable fits of the spectra.
However, once the background is subtracted, one obtains a clear picture of the effect
of Al doping on the far infrared spectrum of Mg$_{1-x}$Al$_{x}$B$_{2}$, as shown Fig. 2b. 
The x=0 spectrum (see Fig. 2b) can be
described by considering a broad peak centered around 460 cm$^{-1}$,
accompanied by a very broad band around 600 cm$^{-1}$. Since the infrared
active E$_{1u}$ and A$_{2u}$ phonons are predicted around 330 and 400 
cm$^{-1}$,\cite{kortus,kong,bohnen,kunc,satta,yildrim} the two observed bands 
might be ascribed to peaks in the phonon density of states,\cite{osborn} which become 
infrared active due to disorder, as noted above. Our MgB$_{2}$
spectrum is qualitatively in agreement with that previously reported, where a broad
peak centered around 480 cm$^{-1}$ is accompanied by further components at
higher frequencies.\cite{sundar} Our data (see Fig. 2b) show that the 460 
cm$^{-1}$ peak becomes more evident with increasing the Al content. For x$>$0.17
a new absorption peak appears around 700 cm$^{-1}$ and strongly increases with further
increasing doping. Both Raman and infrared spectra thus give evidence that
the Al doping induces substantial modifications of the MgB$_{2}$ optical
properties in the high-doping region. In particular, the appearance of new
high-frequency contributions at high doping indicates remarkable structural
changes.

A quantitative description of the Al doping effects can be achieved by analyzing
the best-fit parameters of the Raman spectra. 
The $\nu _{1}$ peak does not significantly vary with doping, its peak frequency and width
being almost constant in the 0-0.5 x range ($\nu _{1}\simeq $415 cm$^{-1}$, 
$\Gamma _{1}\simeq $100 cm$^{-1}$). More interesting information can be
extracted from the x dependence of the peaks at higher frequencies. In Fig.
3 we report the best fit parameters $\nu _{i}$ and $\Gamma _{i}$ (i=2, 3, 4)
as a function of x. In discussing these results, we stress that
two different structural phases have been observed in x-ray diffraction 
measurements:\cite{slusky} a low-doping (LD)
phase in the 0$<$x$<$0.10 range and a high-doping (HD) phase in the x$>$0.25
range. The two phases are structurally incompatible and a two phase region
(LD + HD) at intermediate x (0.10$<$x$<$0.25) has been proposed.\cite{slusky}
Novel diffraction studies confirm a structural change around 
x$=$0.10,\cite{xiang,bianconi2} and transmission-electron-microscopy observations
revealed for x$>$0.10 the existence of a superstructure, with doubled lattice constant
along the c-axis.\cite{xiang} The existence of three major ordered phases for x$<$0.10,
at x$=$0.30 and at x$=$0.50, and an overall disordered phase for 0.10$<$x$<$0.30
have also been suggested.\cite{bianconi2}

As shown in Fig. 3a, where vertical lines at x$=$0.10 and at x$=$0.25
separate the three regions, $\nu _{2}$ and $\nu _{3}$ are nearly constant in
the stable LD region, which is structurally characterized by small variations
of the lattice parameters.\cite{slusky,xiang} The onset of the HD phase in
the intermediate x-region is marked by the appearance of the new spectral
feature ($\nu _{4}$). When the system enters in the pure HD phase, $\nu _{3}$
and $\nu _{4}$ strongly increase with x, while $\nu _{2}$ remains nearly
constant. The latter result is well consistent with the assignment of the 
$\nu _{2}$ peak to the in-plane E$_{2g}$ stretching mode of the boron atoms,
since the a (in-plane) lattice parameter do not significantly vary with x. 
\cite{slusky,xiang} At the moment, our results do not allow a complete
assignment of the observed Raman lines, possible through a polarization analysis of 
Raman spectra from high-quality single crystals. We just note that the increase 
of the $\nu _{3}$ and $\nu _{4}$ values, simultaneous to the remarkable 
compression of the c (out-of-plane) lattice parameter\cite{slusky} when x 
increases in the HD phase, suggests that the $\nu_{3}$ and $\nu_{4}$ peaks 
could be ascribed to out-of-plane vibrations. 

As concerning the widths $\Gamma _{i}$ reported in Fig.3b, their values are
affected by large errors in the LD phase owing to the nearly unstructured
Raman spectra (see Fig.1). Bearing in mind the importance of the e-ph interaction 
on the width of the $\nu _{2}$ (E$_{2g}$) phonon, the most interesting information 
is the x dependence of $\Gamma _{2}$. As a matter of fact, consistently with the
observed suppression of superconductivity,\cite{slusky,xiang,bianconi2} 
$\Gamma _{2}$ and thus the el-ph interaction strongly decreases when x increases
in the HD phase. The overall decrease of $\Gamma _{3}$ and $\Gamma _{4}$ may
also reflect the decrease of the e-ph interaction with increasing x.

Although a comparison among the absolute intensities of different spectra may
be questionable, a comparative analysis of the peak integrated
intensities $I_{i}$ can be safely performed. In Fig. 4a the $I_{2}/I_{3}$ and 
$I_{2}/I_{4}$ values as a function of x are reported. It is well evident that
both the ratios decrease for x$>$0.25, albeit $I_{2}/I_{4}$ is much steeper
than $I_{2}/I_{3}$. This result clearly reflects the progressive transformation 
of the structure from the LD to the HD phase. 
The comparison between the x-dependence of T$_{c}$ reported in Fig. 4b \cite
{bianconi2} and the intensity ratios reported in Fig. 4a clearly shows the
correlation between superconductivity and the phonon structures, i.e. the structural
properties of Mg$_{1-x}$Al$_{x}$B$_{2}$. The relevance of the $\nu _{2}$
mode with respect to the $\nu _{3}$ and in particular 
to the $\nu _{4}$ mode (characteristic of the HD phase) clearly
decreases with increasing x in the HD phase, and conversely the transition
to the superconducting state tends to be inhibited.

In conclusion, the measured Raman and infrared spectra give clear evidence of the
structural difference between the low-doping and the high-doping phases of 
Mg$_{1-x}$Al$_{x}$B$_{2}$.\cite{slusky,xiang,bianconi2} 
The x-dependence of the $\nu_{2}$ peak supports its assignment to the E$_{2g}$ mode. 
Therefore, the remarkable narrowing of the $\nu_{2}$ phonon observed for x$>$0.25, implying 
the decrease of the e-ph interaction, can be related to the suppression of superconductivity.
Moreover, the decrease of the critical temperature can be directly related to the decrease 
of the relative intensity of the $\nu _{2}$ (E$_{2g}$) mode with respect to 
the $\nu_{3}$ and $\ \nu_{4}\ $ modes which dominate the high-doping phase.

\bigskip\
{\bf Figure captions}

Fig. 1- a) Raman spectrum of MgB$_{2}$ and the best fit curve from Eq.1
(solid line). The three phonon contributions (solid lines) and the 
$\nu = 0$ background (dashed line) are also shown separately. b) The Raman
spectra of Mg$_{1-x}$Al$_{x}$B$_{2}$ at x$=$0, 0.08, 0.17, 0.25, 0.33, 0.41,
0.50 after background subtraction. The spectra are shifted vertically for clarity.

\smallskip\ 

Fig. 2 -a) Far infrared optical density of MgB$_{2}$. 
The intense background is very evident (dashed line). b) Optical densities 
of Mg$_{1-x}$Al$_{x}$B$_{2}$ at x$=$0, 0.17, 0.33, 0.41, 0.45 after
background subtraction. The spectra are shifted vertically for clarity.

\smallskip\ 

Fig. 3 - Best fit values of $\nu _{i}$ (a) and $\Gamma _{i}$ (b) (i=2,3,4) as
function of x from the analysis of the Raman spectra. Vertical dashed lines indicate
the x=0.10 and x=0.25 values.

\smallskip\ 

Fig. 4 - a) Intensity ratios of the peak integrated intensities $I_{2}/I_{3}$ 
and $I_{2}/I_{4}$ as a function of x from the analysis of Raman spectra. b) T$_{c}$ as a 
function of x from Ref.16. The vertical dashed lines indicate 
the x=0.25 value. 

\bigskip\

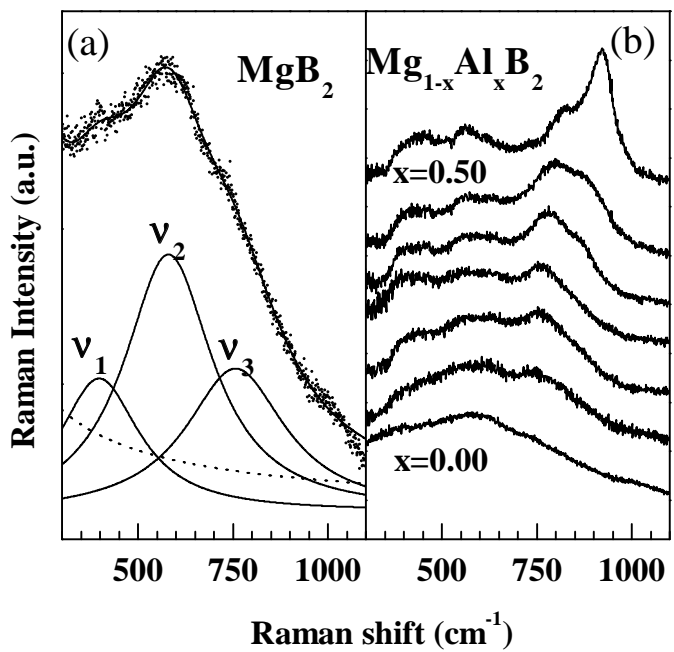
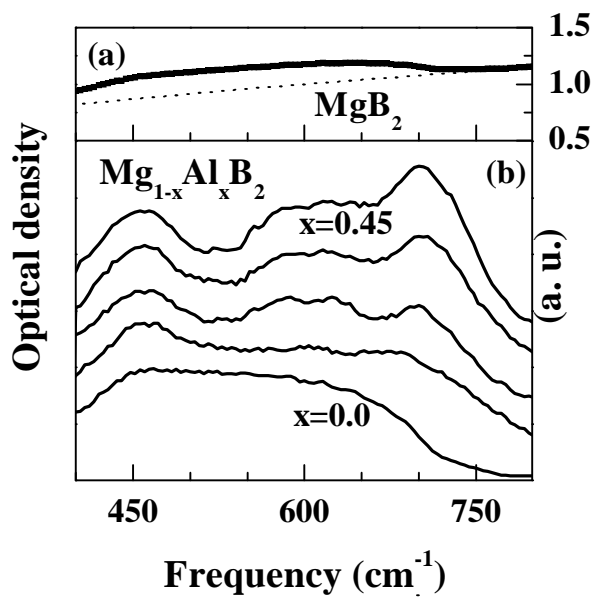
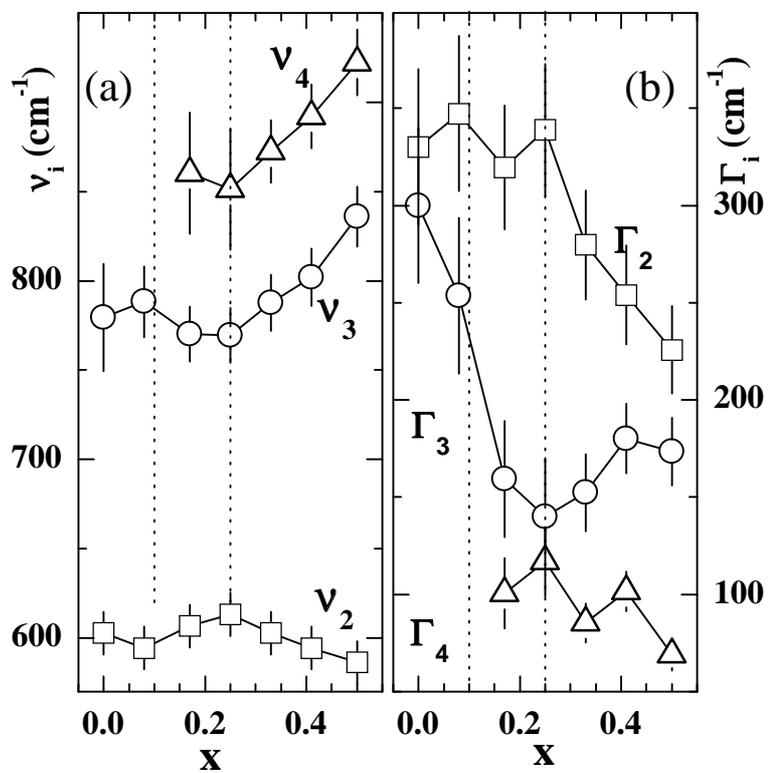
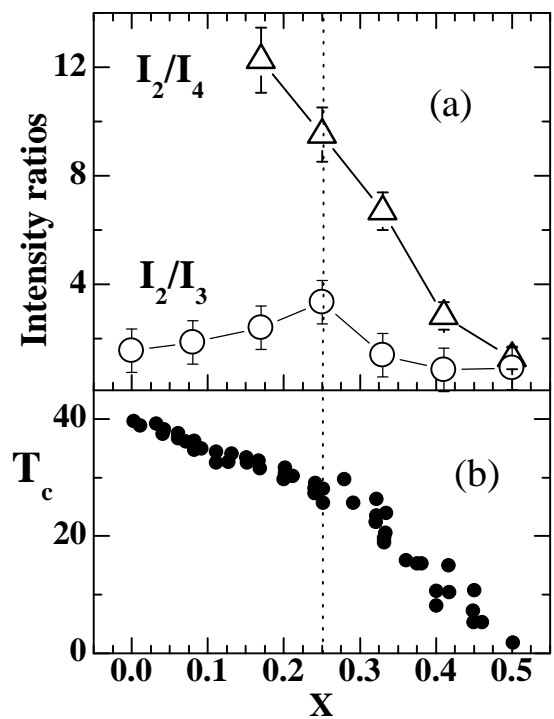
\end{document}